\documentclass[twocolumn,amsmath,amssymb]{revtex4}
\usepackage{graphicx}
\usepackage{dcolumn}
\usepackage{bm}
\begin{document}

\title{Fractal butterflies in buckled graphene-like materials}

\author{Vadym M. Apalkov}
\affiliation{Department of Physics and Astronomy, Georgia State University,
Atlanta, Georgia 30303, USA}
\author{Tapash Chakraborty$^\ddag$}
\affiliation{Department of Physics and Astronomy,
University of Manitoba, Winnipeg, Canada R3T 2N2}

\date{\today}
\begin{abstract}
We study theoretically the properties of buckled graphene-like materials, such as silicene 
and germanene, in a strong perpendicular magnetic field and a periodic potential. We analyze 
how the spin-orbit interaction and the perpendicular electric field influences the energy 
spectra of these systems. When the magnetic flux through a unit cell of the periodic potential 
measured in magnetic flux quantum is a rational number, $\alpha = p/q$, then in each Landau 
level the energy spectra have a band structure, which is characterized by the corresponding 
gaps. We study the dependence of those gaps on the parameters of the buckled graphene-like 
materials. Although some gaps have weak dependence on the magnitude of the spin-orbit 
coupling and the external electric field, there are gaps that show strong nonomonotic 
dependence on these parameters. For $\alpha=1/2$, the spin-orbit interaction also 
opens up a gap at one of the Landau levels. The magnitude of the gap increases with 
spin-orbit coupling and decreases with the applied electric field. 
\end{abstract}
\maketitle

\section{Introduction}

Electron dynamics in a periodic potential subjected to a perpendicular magnetic field was 
the subject of intensive theoretical and experimental research for several decades
\cite{review,harper,langbein,hofstadter,early_work,rhim_park,cuniberti,bistritzer}. That 
interest on this particular subject derives from its unique energy spectrum, which as a 
function of the magnetic flux per lattice cell of the periodic potential reveals a fractal 
pattern, known as the Hofstadter's butterfly (due to the pattern resembling the butterflies). 
This is the first example of fractal pattern realized in a quantum system. The parameter 
which determines the fractal structure of the energy spectrum is the magnetic flux 
$\tilde{\alpha}$ through a unit cell measured in units of the magnetic flux quantum. In 
the case of a weak magnetic field $B$, first the periodic potential forms the Bloch bands 
and then the external magnetic field splits each Bloch band of the periodic potential 
into minibands of Landau level (LL) type, the number of which is determined by the 
parameter $\tilde{\alpha}$. In the strong magnetic field regime, the relevant parameter 
is $\alpha = 1/\tilde{\alpha}$ which is zero when $B\rightarrow \infty$. In this case, 
the energy spectra of an electron can be described by the formation of the LL spectrum 
and then splitting of the LL states into minibands by the periodic potential. Here the 
number of minibands is determined by the parameter $\alpha$, i.e., for rational $\alpha 
= p/q$ where $p$ and $q$ are integers, within a single LL there are $q$ minibands with 
$p$ degeneracy. To observe the fractal pattern in a reasonable range of the magnetic 
field, the important requirement is that the lattice structure has a large period. 
This can be achieved in artificial superlattices based on semiconductor nanostructures 
\cite{early_work}. Observation of a clear fractal pattern remained a challenge however. 
Finally, in 2013 the Hofstadter butterfly structure with clear fractal pattern was 
reported \cite{dean_13,hunt_13,geim_13} in monolayer and bilayer graphene \cite{iet}. 
In these experiments, the periodic lattice with a period $\approx 10$ nm was created 
naturally by the moire pattern that appears when graphene is placed on hexagonal boron 
nitride with a twist \cite{decker,Xue}. In addition to the natural formation of the 
period structure in graphene on boron nitride substrate, that system has also the 
unique relativistic energy dispersion, which results in special type of 
Landau levels \cite{graphene_book,abergeletal,chapter-8}.

Recently discovered new Dirac-type materials, such as silicene and germanene 
\cite{kara,tang,rostami,kormnyos,takeda,yao,liu_11,wang_12,zheng_12} bring in 
additional remarkable features and additional control parameters into the structure 
of the Dirac energy spectrum.  These materials are monolayers of silicon and 
germanium with hexagonal lattice structures where the low energy charge carriers 
are also massless Dirac fermions \cite{lalmi,padova,vogt,chen_12,chun-liang,feng,fleurence} 
just as in graphene \cite{graphene_book,abergeletal}. Experimentally, the two-dimensional 
(2D) silicene was synthesized on Ag(111) \cite{vogt,chen_12,chun-liang,feng} and 
zirconium diboride substrates \cite{fleurence}, while germanene was grown on Ag 
\cite{dvila,li_14} and Pt \cite{li_14} substrates. The main difference between 
silicene/germanene and graphene is that due to the larger radius of the Si/Ge 
atom compared to the C atom, the corresponding hexagon lattices in germanene and 
silicene have {\it buckled} structure \cite{meng}, i.e., the two sublattices 
(say $A$ and $B$) in these systems are displaced vertically by a finite distance 
$L^{}_z $. As a result, silicene and germanene have large spin-orbit interactions, 
which opens the band gaps at the Dirac points ($\Delta^{}_{\rm so} \approx 1.55 - 
7.9$ meV for silicene \cite{yao,liu} and $\Delta^{}_{\rm so} \approx 24 - 93$ meV 
for germanene \cite{yao,liu}). In the case of graphene, in contrast, the corresponding 
spin-orbit-induced gap is tiny, 25 $\mu$eV \cite{gmitra}. The buckled structure of 
silicene/germanene lattice also allows for the band gap to be controlled by an 
applied perpendicular electric field \cite{motohiko} and the size of the band gap 
increases almost linearly with the electric field. These properties have important 
implications for a highly correlated electron systems in these graphene-like but 
novel systems \cite{buckled}, in particular, in the fractional quantum Hall effect 
regime \cite{chapter-8,fqhe}. The buckled graphene-like materials with a strong 
spin-orbit interaction and sensitivity to the external electric field significantly 
modify the energy spectrum of the system, which should be also visible in the 
Hofstadter's butterfly pattern of these materials. Below we study the properties of 
the buckled graphene-like materials, silicene and germanene, 
placed in a perpendicular magnetic field and a periodic potential. 
 
\begin{figure}
\begin{center}
\includegraphics[width=8cm]{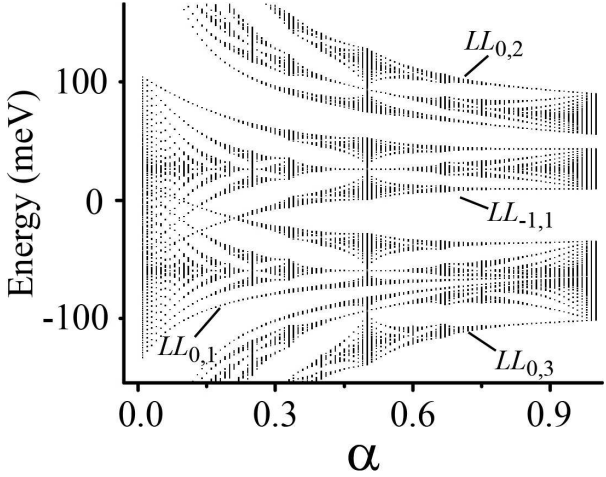}
\end{center}
\caption{Energy spectrum (Hofstadter butterfly) of germanene monolayer as a function 
of the parameter $\alpha$, which is the inverse magnetic flux through the unit cell 
in units of the flux quantum. The external electric field is $E^{}_z= 50 $ mV/\AA. 
The amplitude of the periodic potential is $V^{}_0 = 40$ meV and its period is 
$a = 20$ nm.
}
\label{Fig1_germanene_Q20_V40_E50}
\end{figure}

\section{Theoretical framework}

We consider an electron in a silicene/germanene monolayer in an external perpendicular 
magnetic field and a periodic potential. The Hamiltonian, ${\cal H}$, of such an 
electron consists of the kinetic energy term ${\cal H}^{}_0$ and the periodic potential 
$V(x,y)$,
\begin{equation}
{\cal H} = {\cal H}^{}_0  + V(x,y).
\label{Hgeneral}
\end{equation}
In a magnetic field $B$, the kinetic energy part of the Hamiltonian has the following 
matrix form 
\cite{ezawa}
\begin{equation}
{\cal H}^{}_0 = \left( 
\begin{array}{cccc}
\Delta^{}_{+}(E^{}_z)   &  \hbar \omega^{}_c \hat{a}  & i \frac{\sqrt{2} \hbar a
\lambda^{}_{\rm R}}{\ell^{}_0} \hat{a}^+ & 0 \\
\hbar \omega^{}_c \hat{a}^+ &  - \Delta^{}_+ (E^{}_z)  & 0 & - i \frac{\sqrt{2} \hbar a
\lambda^{}_{\rm R}}{\ell^{}_0} \hat{a}^+ \\
-i \frac{\sqrt{2} \hbar a \lambda^{}_{\rm R}}{\ell^{}_0} \hat{a} & 0 & \Delta^{}_- 
(E^{}_z) & \hbar \omega^{}_c \hat{a} \\
0 & i \frac{\sqrt{2} \hbar a \lambda^{}_{\rm R}}{\ell^{}_0} \hat{a} & \hbar \omega^{}_c 
\hat{a}^+ & - \Delta^{}_- (E^{}_z)
\end{array}
\right),
\label{H0}
\end{equation}
where $\lambda^{}_{\rm R}$ is the Rashba spin-orbit constant, $\ell^{}_0 = \sqrt{\hbar 
c/eB}$ is the magnetic length, $\omega^{}_c = \sqrt{2}\hbar v^{}_{\rm F} /\ell^{}_0$, 
$v^{}_{\rm F}$ is the Fermi velocity, $\Delta^{}_{\pm } = \mp \lambda^{}_{\rm SO} + 
L^{}_z E^{}_z$. Here $\lambda^{}_{\rm SO}$ is the spin-orbit constant, $2L^{}_z$ 
is the separation of two sublattices $A$ and $B$ in the $z$ direction, and $E^{}_z$ 
is the external perpendicular electric field. For germanene and silicene, the 
parameters in the above Hamiltonian are $v^{}_{\rm F} = 7.26\times 10^5$ m/s, 
$L^{}_z = 0.33$ \AA, $\lambda^{}_{\rm SO} = 43$ meV, $\lambda ^{}_{\rm R}=10.7$ meV 
for germanene and $v^{}_{\rm F} = 8.47\times 10^5$ m/s, $L^{}_z = 0.23$ \AA, 
$\lambda^{}_{\rm SO} = 3.9$ meV, $\lambda ^{}_{\rm R}=0.7$ meV  for silicene. 

\begin{figure}
\begin{center}
\includegraphics[width=8cm]{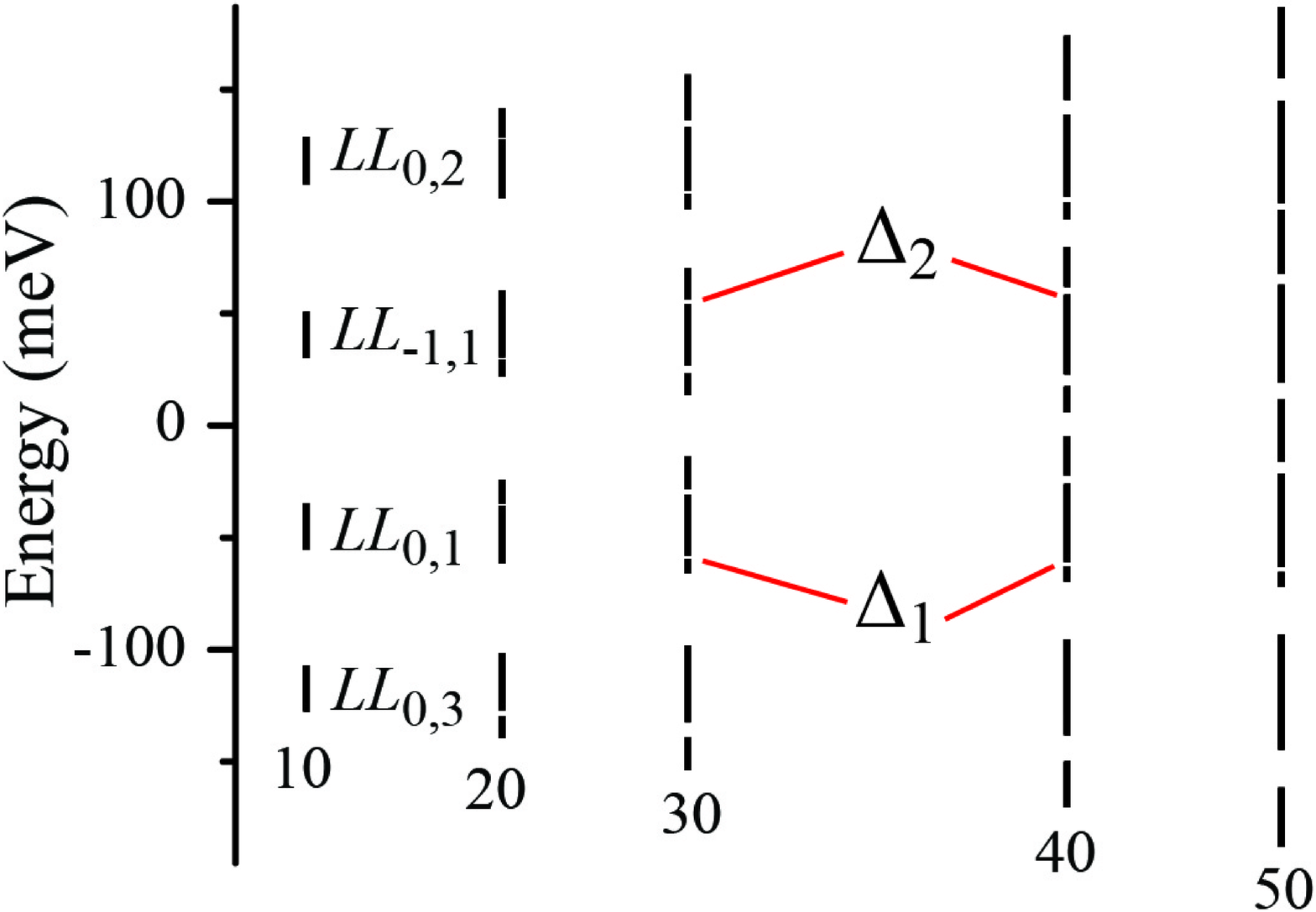}
\end{center}
\caption{Energy spectra of germanene monolayer for $\alpha =1/3$. The external 
electric field is $E^{}_z= 10 $ mV/\AA. The numbers near the lines are the amplitudes 
of the periodic potential, $V^{}_0$. The period of the potential is $a = 20$ nm. 
Two gaps $\Delta^{}_1$ and $\Delta^{}_2$ are marked in the figure.
}
\label{Fig_Ez10}
\end{figure}

The wavefunctions of the Hamiltonian (\ref{H0}) have four components and can be 
expressed in terms of the nonrelativistic wave functions $\phi^{}_{n,k}$, which 
correspond to $n$-th conventional nonrelativistic Landau level and have the 
in-plane $y$ component of the wave vector $k$, 
\begin{equation}
\left| n k\right \rangle  = \left( C^{}_1 \phi^{}_{n,k}, C^{}_2 \phi^{}_{n+1,k}, C^{}_3
\phi^{}_{n-1,k}, C^{}_4 \phi^{}_{n,k} \right).
\end{equation}
In this basis the Hamiltonian (\ref{H0}) takes the form 
\begin{equation}
{\cal H}_0^{(n)} = \left(  
\begin{array}{cccc}
\Delta^{}_{+}(E^{}_z)   &  \kappa^{}_{n+1}  & i \alpha ^{}_n & 0 \\
\kappa^{}_{n+1}  &  - \Delta^{}_+ (E^{}_z)  & 0 & - i \alpha^{}_{n+1} \\
-i \alpha^{}_n & 0 & \Delta^{}_- (E^{}_z) & \kappa^{}_n \\
0 & i \alpha^{}_{n+1} & \kappa^{}_n & - \Delta^{}_- (E^{}_z)
\end{array}
\right),
\label{H0m}
\end{equation}
where $\kappa^{}_n = \hbar \omega^{}_c \sqrt{n}$, $\alpha^{}_ n = \sqrt{2n} \hbar a
\lambda^{}_{\rm R}/\ell^{}_0$
and the wave functions are determined now by the coefficients $(C^{}_1, C^{}_2, 
C^{}_3, C^{}_4)$. 

The conventional Landau wave functions $\phi^{}_{n,k}$ have the form
\begin{equation}
\phi^{}_{n,k} (x,y) = \frac{e^{i k y}}{\sqrt{L}} 
\frac{e^{-(x-x^{}_k )^2/2\ell_0^2}}{\sqrt{ \pi^{1/2} \ell^{}_0 2^n n!}}
 H^{}_n (x-x^{}_k),
\label{PhiK}
\end{equation}
where $L$ is the length of a sample in the $y$ direction, $k$ is the $y$ component 
of the electron wave vector,  $x^{}_k=k\ell^{2}_0$, and $H^{}_n(x)$ are the Hermite 
polynomials. 

Depending on the value of $n$, the energy spectrum of the Hamiltonian (\ref{H0m}) 
has the following properties: (i) for $n=-1$ there is only one Landau level, which 
is characterized by the wavefunction 
\begin{equation}
\left| -1, k\right \rangle = (0,\phi^{}_{0,k},0,0)
\end{equation}
and has the energy of $E^{}_{n=-1} = - \Delta^{}_+ (E^{}_z)$; (ii) for $n=0$ there 
are three different Landau levels,
\begin{equation}
\left| 0, k\right \rangle  = (C^{}_1\phi^{}_{0,k} ,C^{}_2\phi^{}_{1,k},0,C^{}_4
\phi^{}_{0,k} )
\end{equation}
 and their energies are determined by the following cubic equation 
\begin{eqnarray}
& & E^3  + \Delta^{}_- E^2 - \left( \Delta_+^2  + \lambda^2_1 +\kappa_1^2 
\right) E  \nonumber \\
& & + \left( \lambda_1^2 \Delta^{}_+ - \Delta_+^2 \Delta^{}_- - \kappa_1^2 
\Delta^{}_-  \right) = 0;
\label{eigen1}
\end{eqnarray}
In the absence of the external electric field ($E^{}_z =0$), one of the solutions 
of Eq.\ (\ref{eigen1}) is $E^{}_0 = \Delta^{}_+ = -\lambda^{}_{\rm SO}$. The 
corresponding wavefunction is
\begin{equation}
\left| 0, k\right \rangle = (i \alpha^{}_1\phi^{}_{0,k}, 0, 0, \kappa^{}_1\phi^{}_{0),k})
\end{equation}
and (iii) for $n>0$ there are four different Landau levels with the general 
structure $\left( C^{}_1 \phi^{}_{n,k}, C^{}_2 \phi^{}_{n+1,k}, C^{}_3 \phi^{}_{n-1,k}, 
C^{}_4 \phi^{}_{n,k} \right)$. In what follows, we mainly consider the properties of 
the $n=-1$ and $n=0$ Landau levels. 

\begin{figure}
\begin{center}
\includegraphics[width=8cm]{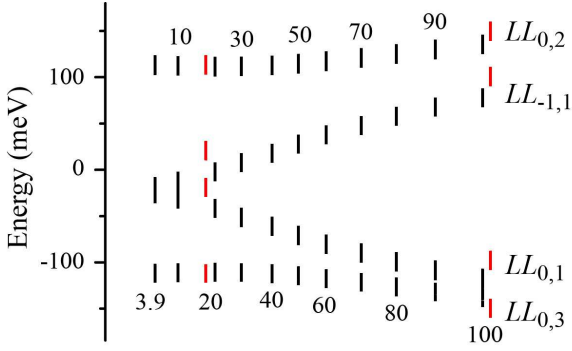}
\end{center}
\caption{Energy spectra of buckled graphene-like materials for $\alpha=1/3$. The 
external electric field is $E^{}_z= 100 $ mV/\AA\, (black lines) and $E^{}_z= 0$ 
(red lines). The numbers near the lines are the SO interaction constants, 
$\lambda^{}_{\rm SO}$. The period of the potential is $a = 20$
nm and its amplitude is $V^{}_0 = 20$ meV.
}
\label{Fig_SO}
\end{figure}

The periodic potential, $V(x,y)$, in the Hamiltonian (\ref{Hgeneral}) is characterized 
by its period $a^{}_0$ and amplitude $V^{}_0$. We assume that the potential has 
the following profile 
\begin{equation}
V(x,y) = V^{}_0 \left[\cos(q^{}_x x)+\cos(q^{}_y y)\right],
\label{Vxy}
\end{equation}
where  $q^{}_x = q^{}_y = q^{}_0 = 2\pi/a^{}_0$.
The periodic potential mixes the electron states $\Psi^{}_{n,k}$ within a single LL, 
i.e., states with the same value of the LL index $n$ and different values of $k$, 
and also mixes the states of different LLs. The strength of this mixing is determined 
by the matrix elements of the periodic potential $V(x,y)$ between 
the LL states.

\begin{figure}
\begin{center}
\includegraphics[width=5cm]{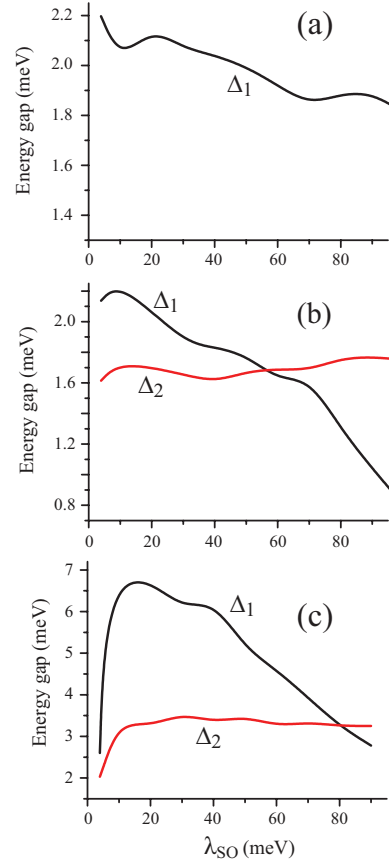}
\end{center}
\caption{Gaps $\Delta^{}_1$ and $\Delta^{}_2$, defined in Fig.\ \ref{Fig_Ez10}, 
as functions of the SO interaction, $\lambda^{}_{\rm SO}$. The period of the 
potential is 20 nm. (a) The electric field is zero and the amplitude of periodic 
potential is 10 meV. (b)  The electric field is 100 mV/\AA\, and the amplitude 
of the periodic potential is 10 meV. (c) The electric field is
100 mV/\AA\, and the amplitude of the periodic potential is 20 meV.
}
\label{Fig_gap}
\end{figure}

The matrix elements of the periodic potential in the basis of the LL wave functions 
of the buckled materials have the following form  
\begin{eqnarray}
& & \left\langle n^{\prime }k^{\prime}\right|\cos(q^{}_0 y)\left| n k\right 
\rangle = \nonumber \\  
& & \frac{i^{n-n^{\prime}}}{2}  
\left\{\delta^{}_{k^{\prime}, k+q^{}_0} 
  +  (-1)^{n-n^{\prime }} \delta^{}_{k^{\prime}, k-q^{}_0} 
  \right\}      \nonumber \\
 & &  \times 
\left[C^{}_{n,1} C^{}_{n^{\prime},1} M^{}_{|n^{\prime}|-1,|n|-1} 
+ C^{}_{n,4} C^{}_{n^{\prime},4} M^{}_{|n^{\prime}|+1,|n|+1}  \right. \nonumber \\
& & + \left. \left( C^{}_{n,2} C^{}_{n^{\prime},2}  + C^{}_{n,3} C^{}_{n^{\prime},3}\right) 
M^{}_{|n^{\prime}|,|n|} 
     \right] 
\label{Vy}
\end{eqnarray}
and 
\begin{eqnarray}
& & \left\langle n^{\prime}k^{\prime}\right|\cos(q^{}_0 x) \left| n k\right
\rangle = \frac{\delta^{}_{k^{\prime}, k}}{2}  \left[e^{i q^{}_0 k\ell_0^2} +(-1)^{n-n^{\prime}}
e^{-i q^{}_0 k\ell_0^2} \right]  
\nonumber \\
& &\times 
\left[C^{}_{n,1} C^{}_{n^{\prime},1} M^{}_{|n^{\prime}|-1,|n|-1} 
+ C^{}_{n,4} C^{}_{n^{\prime},4} M^{}_{|n^{\prime}|+1,|n|+1}  \right. \nonumber \\
& & + \left. \left( C^{}_{n,2} C^{}_{n^{\prime},2}  + C^{}_{n,3} C^{}_{n^{\prime},3}\right) 
M^{}_{|n^{\prime}|,|n|} 
     \right],
 \label{Vx}
\end{eqnarray}
where 
\begin{equation}
M^{}_{n^{\prime},n} = \left(\frac{m!}{M!}\right)^{1/2} e^{-Q/2} Q^{|n^{\prime}
-n|/2} L_m^{|n^{\prime }-n|} (Q),
\end{equation}
and $Q = q_0^2 \ell_0^2/2$, $m=\min (n^{\prime},n)$, $M=\max (n^{\prime},n)$.

We study the regime of the strong magnetic field and consider the basis of four LLs, 
which correspond to $n=-1$ and $n=0$. Within these basis we construct the Hamiltonian 
matrix taking into account the matrix elements (\ref{Vy}) and (\ref{Vx}) of the periodic 
potential, and evaluate the corresponding energy spectrum as a function of parameter 
$\alpha = Ba^2_0 /\Phi^{}_0$, where $\Phi^{}_0 = h/(2e)$ is the 
magnetic flux quantum. 

\begin{figure}
\begin{center}
\includegraphics[width=8cm]{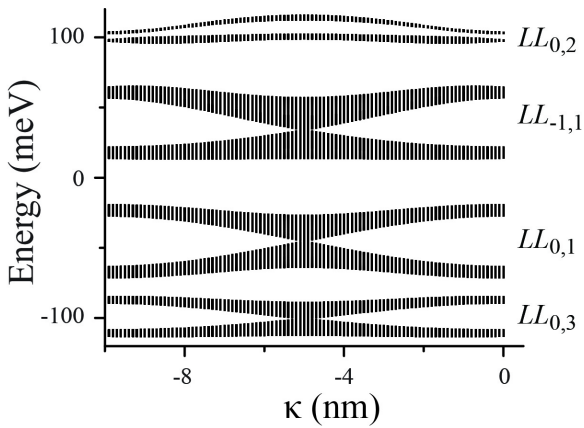}
\end{center}
\caption{Energy spectrum of the germanene monolayer as a function of the effective 
wave vector $\kappa$. The external electric field is zero. The amplitude of periodic 
potential is 40 meV and its period is 20 nm.
}
\label{Fig_spectrum_1_2}
\end{figure}

\section{Numerical Results and Discussion}

\subsection{The butterfly Energy Spectrum}

The energy spectra of buckled graphene-like materials in the regime of the strong 
magnetic field, i.e., weak periodic potential, is shown in 
Fig.\ \ref{Fig1_germanene_Q20_V40_E50} as a function of the parameter $\alpha$, 
which changes from 0 to 1. We consider only four LLs with the label scheme shown 
in the figure, i.e., $LL^{}_{-1,1}$ corresponds to the LL with index $n=-1$ 
(there is only one LL with that index), while $LL^{}_{0,1}$, $LL^{}_{0,2}$, and 
$LL^{}_{0,3}$ correspond to three LLs with index $n=0$. The results in 
Fig.\ \ref{Fig1_germanene_Q20_V40_E50} correspond to germanene, which has the 
largest SO interaction constant, $\lambda^{}_{\rm SO} = 43$ meV. The results 
clearly illustrate the formation of the Hofstadter butterfly structures in each LL. 
For small $\alpha $, there is an overlap of the butterfly structures of the 
Landau levels $LL^{}_{0,1}$ and $LL^{}_{-1,1}$. Although the overlap is large, 
the coupling of $LL^{}_{0,1}$ and $LL^{}_{-1,1}$ states is weak. This is due 
to the fact that the overlap of the corresponding wave functions is small. For 
the zero external electric field, this overlap is exactly zero. For $\alpha > 0.7$, 
i.e., in a weak magnetic field there is an overlap of the $LL^{}_{0,1}$ and 
$LL^{}_{0,2}$ Landau levels. In this case the overlap results in a relatively 
strong coupling of the corresponding states and modification of the energy spectrum. 

The results shown in Fig.\ \ref{Fig1_germanene_Q20_V40_E50} illustrate the typical 
structure of the energy spectra of buckled graphene-like materials. Variation of 
the external electric field, $E^{}_z$, and the strength of the SO interaction, 
$\lambda^{}_{\rm SO}$, change the energy spectra of the system. These changes 
can be described by studying the values of the energy gaps for rational $\alpha 
=p/q$, where $q-1$ gaps exist in each LL. For example, for $\alpha = 1/3$ there 
are three bands and correspondingly two gaps in each LL, while for $\alpha =1/2$ 
there are two bands and one gap in each LL. In the case of overlapping LLs the 
gaps can also be closed. Below we study the properties of the energy spectra of 
buckled graphene-like materials for $\alpha = 1/3$ and $\alpha = 1/2$. 

\subsection{$\alpha = 1/3$}

For $\alpha =1/3$ there are three bands in each LL with the corresponding two 
gaps. In Fig.\ \ref{Fig_Ez10} the energy spectra for $\alpha = 1/3$ are shown 
for four LLs and different amplitudes of the periodic potential, $V^{}_0$. In 
each LL the three bands are clearly visible. With increasing $V^{}_0$ the widths 
of the bands and correspondingly the band gaps increase. For large $V^{}_0$ 
the coupling of the states of different LLs modifies the band structure of the 
LLs. We characterize this effect by studying the magnitudes of the two gaps marked 
in Fig.\ \ref{Fig_Ez10} by $\Delta^{}_1$, which is the higher energy gap in the 
Landau level $LL^{}_{-1,1}$, and $\Delta^{}_2$, which is the lower energy gap in
the Landau level $LL^{}_{0,1}$.
 
\begin{figure}
\begin{center}
\includegraphics[width=5cm]{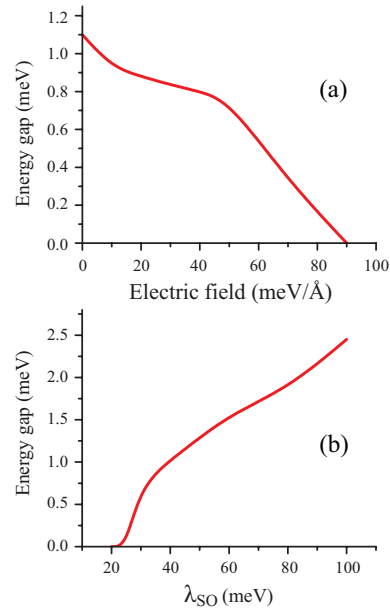}
\end{center}
\caption{Energy gap $\Delta $ between two bands in $LL^{}_{0,2}$. (a) The period 
of the potential is 20 nm and its amplitude is 40 meV. The SO coupling is 
$\lambda^{}_{\rm SO} = 43$ meV. (b) The electric field is
$10$ mV/\AA. The period of the potential is 20 nm and its amplitude is 40 meV.
}
\label{Fig_gap1_2}
\end{figure}

The effect of the SO interaction strength on the arrangement of the LL bands is 
illustrated in Fig.\ \ref{Fig_SO}. For small $\lambda^{}_{\rm SO}$ ($\lesssim 
20$ meV) the LLs $LL^{}_{-1,1}$ and $LL^{}_{0,1}$, which are broadened by the 
periodic potential, overlap that can close some gaps within these LLs. This 
corresponds to the case of silicene, for which $\lambda^{}_{\rm SO}=3.9$ meV. 
With increasing $\lambda^{}_{\rm SO}$, the LLs $LL^{}_{-1,1}$ and $LL^{}_{0,1}$ 
become well separated. Compared to other LLs, the energies of the LLs $LL^{}_{-1,1}$ 
and $LL^{}_{0,1}$ have strong dependence on the SO coupling. This strong dependence 
results in the overlap of the LLs $LL^{}_{0,1}$ and $LL^{}_{0,3}$ for a large spin 
orbit coupling and large external electric field. The overlap of the LLs $LL^{}_{0,1}$ 
and $LL^{}_{0,3}$ is visible in Fig.\ \ref{Fig_SO} for $\lambda^{}_{\rm SO} =100$ 
meV and the electric field $E^{}_z = 100$ mV/\AA, while for $E^{}_z = 0$ there 
is no overlap of the LLs even for large $\lambda^{}_{\rm SO}$. In Fig.\ \ref{Fig_gap}, 
the dependence of the energy gaps $\Delta^{}_1$ and $\Delta^{}_2$, which are defined 
in Fig.\ \ref{Fig_Ez10}, on the SO coupling, $\lambda^{}_{\rm SO}$. For zero electric 
field [Fig.\ \ref{Fig_Ez10}(a)] the gaps $\Delta^{}_1$ and $\Delta^{}_2$ are 
exactly the same. In this case the dependence of $\Delta^{}_1$ on the SO interaction 
strength is relatively weak. The gap $\Delta^{}_1$ changes by only $\approx 15$\% 
when the coupling $\lambda^{}_{\rm SO}$ increases from 3 meV (silicene) to 100 meV. 

The dependence of the gaps $\Delta^{}_1$ and $\Delta^{}_2$ on $\lambda^{}_{\rm SO}$ 
becomes more pronounced for large electric fields. The gaps $\Delta^{}_1$ and 
$\Delta^{}_2$ in this case are not equal. The gap $\Delta^{}_2$ still has a very 
weak dependence on $\lambda^{}_{\rm SO}$. This gap corresponds to the high energy 
gap for $LL^{}_{-1,1}$. At the same time, the gap $\Delta^{}_1$ has a very strong 
dependence on $\lambda^{}_{\rm SO}$ and with increasing $\lambda^{}_{\rm SO}$ it 
is strongly suppressed [Fig.\ \ref{Fig_gap}(b)]. The gap $\Delta^{}_{1}$ changes 
from 2 meV for small $\lambda^{}_{\rm SO}\approx 3$ meV to 0.8 meV for large 
$\lambda^{}_{\rm SO} \approx 100$ meV. This suppression is due to the fact that 
with increasing $\lambda^{}_{\rm SO}$ the separation between the LLs $LL^{}_{0,1}$ 
and $LL^{}_{0,3}$ decreases (see Fig.\ \ref{Fig_SO}), which results in stronger 
coupling of the states of these LLs, and correspondingly a strong change in the 
value of the gap $\Delta^{}_1$. The gap, $\Delta^{}_1$, has also a nonmonotonic 
dependence on $\lambda^{}_{\rm SO}$, which is more pronounced for larger amplitudes 
of the periodic potential, $V^{}_0$. As an example, for $V^{}_0 = 20$ meV (see 
Fig.\ \ref{Fig_gap}(c)), the gap $\Delta^{}_1$ first increases from small value 
of $\approx 2.5$ meV for small $\lambda^{}_{\rm SO}$, then reaches its maximum 
of $\approx 6.5$ meV for $\lambda^{}_{\rm SO} \approx 20$ meV, and finally 
decreases for large $\lambda^{}_{\rm SO}$. 

\subsection{$\alpha = 1/2$}

For $\alpha = 1/2$ there are two bands in each LL. Specific to this case is the 
fact that the gap between these two bands in each LL is zero. A gap can however be 
opened by the Coulomb interaction between electrons \cite{apalkov_14,fractal_interaction}. 
In the case of buckled graphene-like materials there are other parameters, the SO 
coupling and the external electric field, that can modify the band structure of the
LLs. To characterize the band structure of the LLs, we introduce an ``effective 2D 
wave vector", $(\kappa^{}_x,\kappa^{}_y)$, which reflects the periodicity of the 
wave functions in the reciprocal space. This periodicity follows from the expressions 
of the matrix elements of the periodic potential [Eqs.\ (\ref{Vy})-(\ref{Vx})]. 
The effective wave vector $(\kappa^{}_x,\kappa^{}_y)$ has the units of length and 
are defined within an effective Brillouin zone. In Fig.\ \ref{Fig_spectrum_1_2} 
the typical energy spectrum of the germanene layer for $\alpha = 1/2$ is shown 
as a function of $\kappa^{}_x$ and different values of $\kappa^{}_y$. The spectrum 
clearly shows two bands in each LL. The gaps between the bands are zero for the 
LLs $LL^{}_{-1,1}$, $LL^{}_{0,1}$, and $LL^{}_{0,3}$, while there is a finite 
gap $\Delta$ for $LL^{}_{0,2}$. In Fig.\ \ref{Fig_gap1_2} the gap $\Delta$ is 
shown as a function of the electric field and the SO coupling. With increasing 
electric field the gap decreases and finally disappears for large fields. As a 
function of the SO interaction the gap increases with $\lambda^{}_{\rm SO}$, which 
illustrates the fact that the gap is due to the strong SO coupling in the system. 
For small SO coupling, $\lambda \lesssim 20$ meV, the gap is zero and it monotonically 
increases with $\lambda^{}_{\rm SO}$ reaching the value of 2.5 meV for $\lambda^{}_{\rm SO} 
= 100$ meV. Therefore, opening of the gap for $\lambda = 1/2$ can be observed 
only in the germanene monolayer which has a large SO coupling, $\lambda \approx 40$ 
meV, while in the silicene monolayer with small SO coupling, $\lambda^{}_{\rm SO} 
\approx 3.4$ meV, the gap is zero.  

\section{Conclusion}

The graphene-like materials, e.g., silicene and germanene, have strong SO 
interaction and the buckled structure that results in these materials being sensitive 
to a perpendicular electric field. In this case the main parameters, which determines 
the unique properties of these materials, are the SO coupling, $\lambda^{}_{\rm SO}$, 
and the external perpendicular electric field, $E^{}_z$. The properties of buckled 
graphene-like materials in a magnetic field and the periodic potential also depend 
on the values of these parameters. That dependence can be described in terms of the 
dependence of the band structure of the LLs for rational values of $\alpha$. One of 
the characteristics of the band structure is the gap between the intra-Landau level 
bands. For $\alpha = 1/3$, some gaps show strong dependence on both the electric field 
and the SO coupling. For a large electric field and large amplitude of the periodic 
potential, the dependence of the gap on the SO coupling is highly nonmonotonic. For 
$\alpha=1/2$, without the SO coupling and for all values of the electric field, all 
gaps are closed. For a large SO interaction, $\lambda^{}_{\rm SO} > 20$ meV, which 
is realized in germanene, the gap in one of the LLs opens. The magnitude of the gap 
increases with $\lambda^{}_{\rm SO}$ and decreases with the electric field. Experimental
confirmation of these predictions will provide a rare peek into the electronic 
properties of these unique materials with emerging properties. Finally, the influence 
of the electron-electron interaction on the butterfly gap structure is an important 
direction for exploration in these and other similar \cite{mos2} Dirac materials, 
that has already seen some progress, theoretically and experimentally, in graphene 
\cite{apalkov_14,fractal_interaction,geim_inter}.

\section{Acknowledgments}
The work has been supported by the Canada Research Chairs Program of the Government 
of Canada.

\end{document}